\documentclass[twocolumn,showpacs,preprintnumbers,amssymb,nofootinbib]{revtex4}

\usepackage{graphicx}

\begin{document}


\title{Magnetic point sources in three dimensional Brans-Dicke  
gravity theories}

\author{\'Oscar J. C. Dias}
\email{oscar@fisica.ist.utl.pt}
\author{Jos\'e P. S. Lemos}
\email{lemos@kelvin.ist.utl.pt}
\affiliation{
Centro Multidisciplinar de Astrof\'{\i}sica - CENTRA, 
Departamento de F\'{\i}sica, Instituto Superior T\'ecnico,
Av. Rovisco Pais 1, 1049-001 Lisboa, Portugal
}%

\date{\today}

\begin{abstract}

We obtain geodesically complete spacetimes generated by  
static and rotating magnetic point sources in an 
Einstein-Maxwell-Dilaton theory 
of the Brans-Dicke type in three dimensions (3D). The theory is 
specified by three fields, the dilaton $\phi$, the graviton 
$g_{\mu\nu}$ and the electromagnetic field $F^{\mu\nu}$, and 
two parameters, the cosmological constant $\Lambda<0$ and the 
Brans-Dicke parameter $\omega$. 
For $\omega=\pm \infty$, our
solution reduces to the magnetic counterpart of the BTZ solution,
while the $\omega=0$ case is equivalent to 4D general
relativity with one Killing vector.
The source for the magnetic field can be interpreted as
composed by a system of two symmetric and superposed electric 
charges. One of the electric charges is at rest and the other 
is spinning.

\end{abstract}


\maketitle
\section{Introduction}

The issue of spacetimes generated by point sources in 
three dimensional (3D) Einstein theory has been object of 
many studies 
(for reviews see \cite{Jackiw_Review}-\cite{Jackiw_book}).
In 1963, Staruszkiewicz \cite{Star} has begun the analysis
of gravity without cosmological constant ($\Lambda=0$) in
3D   coupled to a static massive point source. 
The corresponding space has a conical geometry, i.e., it is 
everywhere flat except at the location of the point source.
The space can be obtained from the Minkoswki space by 
suppressing a wedge and identifying its edges.
The wedge has an opening angle which turns to be proportional 
to the source mass. 
Gott and Alpert \cite{GA} and 
Giddings, Abbot and Kucha\v{r} \cite{GAK}
have studied several features of Einstein gravity in 
3D, 
including the non-existence of the Newtonian limit and the 
presence of a conical geometry.
Deser, Jackiw and t' Hooft \cite{DJH_flat} have generalized 
the analysis of \cite{Star} in order to find the spacetime 
solutions generated by an arbitrary number of  static 
massive point sources. Once more the geometry is conical
with a wedge angle suppressed at each source proportional
to its mass. 
They have also constructed the solution 
corresponding to a massless spinning point source in 
3D Einstein gravity with $\Lambda=0$.
The extension to include massive spinning sources has 
been achieved by Cl\'ement \cite{Clem_spin}.
Their results indicate that (besides the conical geometry
already present in the spinless case) the spacetime can be seen 
as characterized by an helical structure since a complete 
loop around the source ends with a shift in time 
proportional to the angular momentum.

In the context of  Einstein-Maxwell theory but still 
with $\Lambda=0$, Deser and Mazur \cite{Deser_Maz} and
Gott, Simon and Alpert \cite{GSA} have found the solutions
produced by electric charged point sources. 
These spacetimes once more have a conical geometry with an 
helical structure. 
Barrow, Burd and Lancaster \cite{BBL} found the horizonless
spacetime generated by a magnetic charged point source 
in a $\Lambda=0$ background.
Melvin \cite{Melvin} also describes the exterior solution
of electric and magnetic stars in the above theory.

3D static spacetimes generated by open or closed
one-dimensional string sources with or without tension
(in a $\Lambda=0$ background) have
been constructed by Deser and Jackiw \cite{Deser_Jackiw_string}.
The extension to the rotating string source has been done
by Grignani and Lee \cite{Grig_Lee} and by 
Cl\'ement \cite{Clem_string}.
Other exact solutions in 3D gravity theory
produced by extended and stationary sources have been found
by Menotti and Seminara \cite{Men_Sem}.

One of the alternative theories to Einstein gravity in 3D
is the topological massive gravity which includes
the Chern-Simmons term and has been proposed by 
Deser, Jackiw and Templeton \cite{Des_Jac_Temp}.
The spacetimes generated by point sources in this theory
have been obtained by Carlip \cite{Carlip_source} and 
Gerbert \cite{Gerbert_source}.

The spacetimes generated by point sources in 
3D Einstein gravity with non-vanishing
cosmological constant ($\Lambda \neq 0$) have been obtained by 
Deser and Jackiw \cite{DJ_sitter} and by Brown and 
Henneaux \cite{Brown_Hen}. In the de Sitter case 
($\Lambda >0$) there is no one-particle solution.
The simplest solution describes a pair of 
antipodal particles on a sphere with a wedge removed
between poles and with points on its great circle
boundaries identified. In the anti-de Sitter case ($\Lambda<0$),
the simplest solution describes a hyperboloid with 
a wedge removed proportional to the source mass located
at the vertex of the wedge.  
For $\Lambda >0$ there are no other interesting solutions 
with electromagnetic  and other fields, as far as we know.
On the other hand, for $\Lambda <0$ there is a bewildering variety
of solutions. The
black hole solution in 3D Einstein gravity with
$\Lambda <0$ has been found by Ba\~nados, Teitelboim and 
Zanelli \cite{btz_PRL}. The extension to include a radial
electric field has been done by Cl\'ement \cite{CL1} and
Mart\'{\i}nez, Teitelboim and Zanelli \cite{BTZ_Q}
and the inclusion of an azimuthal electric field has made by 
Cataldo \cite{Cat}.
Here, we dedicate a special attention to 
the solutions of an Einstein-Maxwell-Dilaton action of the 
Brans-Dicke type in a $\Lambda <0$ background.
The static uncharged solutions of this theory were found and 
analyzed by S\'a, Kleber and Lemos \cite{Sa_Lemos_Static} and 
the angular momentum has been added by S\'a and Lemos 
\cite{Sa_Lemos_Rotat}.
The uncharged theory is specified by two fields, 
the graviton $g_{\mu\nu}$, the dilaton 
$\phi$, and two parameters, 
the Brans-Dicke parameter $\omega$ and the 
cosmological constant $\Lambda$. 
The pure electrically charged theory is specified by the extra 
electromagnetic field $F^{\mu \nu}$ and was analysed by Dias and Lemos
in \cite{OscarLemos}. This Brans-Dicke theory contains seven different cases 
and each $\omega$ can be viewed as yielding a different dilaton gravity 
theory. For instance, for $\omega=0$ one gets a 
theory related (through dimensional reduction) to 4D 
General Relativity with one Killing vector 
\cite{Lemos,Zanchin_Lemos} and for $\omega=\pm \infty$ one obtains 
3D General Relativity analyzed in 
\cite{btz_PRL,CL1,BTZ_Q}. 
For a review on other 3D  theories, 
specially on black hole solutions, see
\cite{Carlip,OscarLemos}.

In this paper we are interested in  pure magnetic solutions
with $\Lambda<0$. 
The static magnetic counterpart of the BTZ solution has been found
by Cl\'ement \cite{CL1},
Hirschmann and Welch \cite{HW} and Cataldo and Salgado  \cite{Cat_Sal}.
This spacetime generated by a static magnetic point source
is horizonless and reduces to the 3D BTZ black
hole solution of Ba\~nados, Teitelboim and Zanelli 
\cite{btz_PRL} when the magnetic source vanishes.
The extension to include rotation, definition of conserved quantities,
upper bounds for the conserved quantities and an interpretation for
the source of magnetic field has been made 
by Dias and Lemos \cite{OscarLemos_BTZ}.
Kiem and Park \cite{KiemP}, Park and Kim \cite{PKim}
and Koikawa, Maki and Nakamula \cite{KMK} have found
magnetically charged solutions of Einstein-Maxwell-dilaton
theories. 
Many authors have also found self-dual and anti$-$self-dual 
charged solutions of Einstein-Maxwell \cite{ME}, 
Einstein-Maxwell-dilaton and Einstein$-$Maxwell$-$Chern-Simons \cite{MECS} 
theories in 3D (For a complete review see e.g.
\cite{OscarLemos,Cat}).

The aim of this paper is to find and study in detail the static and 
rotating magnetic charged solutions generated by a magnetic
point source in the Einstein-Maxwell-Dilaton action of the 
Brans-Dicke type mentioned above. In this sense it is a follow up
of our previous paper \cite{OscarLemos} on pure electric solutions.
Here, we impose that the only 
non-vanishing component of the vector potential is 
$A_{\varphi}(r)$.  
For the $\omega=\pm \infty$ case, our
solution reduces to the spacetime generated by a magnetic 
point source in 3D Einstein-Maxwell theory with 
$\Lambda<0$, studied in \cite{CL1}, \cite{HW}-\cite{OscarLemos_BTZ}.
The $\omega=0$ case is equivalent to 4D general
relativity with one Killing vector analysed by 
Dias and Lemos \cite{OscarLemos_string}.

The plan of this article is the following. In Section II we set 
up the action and the field equations. The 
static general 
solutions of the field equations are found in section III
and  we analyse in detail the general structure of the 
solutions. 
The angular momentum is added in section IV  and
we calculate the mass, angular momentum, and 
electric charge of the solutions.
In section V we give a physical interpretation for the origin of the
magnetic field source.
Finally, in section VI we present the 
concluding remarks.

\section{FIELD EQUATIONS OF BRANS-DICKE$-$MAXWELL THEORY}

We are going to work with an action of the 
Brans-Dicke$-$Maxwell  type in 3D written in the 
string frame as
\begin{equation}
S=\frac{1}{2\pi} \int d^3x \sqrt{-g} e^{-2\phi}
  {\bigl [} R - 4 \omega ( \partial \phi)^2
           + \Lambda + F^{\mu \nu}F_{\mu \nu}
  {\bigr ]},                                   \label{ACCAO}
\end{equation}
where
$g$ is the determinant of the 3D metric,
$R$ is the curvature scalar,
$F_{\mu\nu} = \partial_\nu A_\mu - \partial_\mu A_\nu$ is the Maxwell tensor,
with $A_\mu$ being the vector potential,
$\phi$ is a scalar field called dilaton, 
 $\omega$ is the 3D Brans-Dicke parameter and
$\Lambda$ is the cosmological constant.
Varying this action with respect to $g^{\mu \nu}$, $A_{\mu}$
and $\phi$ one gets the Einstein, Maxwell and dilaton equations, 
respectively
\begin{eqnarray}
   & &   \frac{1}{2} G_{\mu \nu}
       -2(\omega+1) \nabla_{\mu}\phi \nabla_{\nu}\phi
       +\nabla_{\mu} \nabla_{\nu}\phi
       -g_{\mu \nu} \nabla_{\gamma} \nabla^{\gamma}\phi   \nonumber \\
   & &  \hskip 0.5cm
       +(\omega+2) g_{\mu \nu} \nabla_{\gamma}\phi
               \nabla^{\gamma}\phi-\frac{1}{4}g_{\mu \nu}\Lambda=
              \frac{\pi}{2} T_{\mu \nu},
                                        \label{EQUACAO_MET} \\   
   & &  \nabla_{\nu}(e^{-2 \phi}F^{\mu \nu})=0 \:,
                                     \label{EQUACAO_MAX} \\
   & &  R -4\omega \nabla_{\gamma} \nabla^{\gamma}\phi
          +4\omega \nabla_{\gamma}\phi \nabla^{\gamma}\phi
          +\Lambda =-F^{\gamma \sigma}F_{\gamma \sigma},
                                         \label{EQUACAO_DIL} 
\end{eqnarray}
where $G_{\mu \nu}=R_{\mu \nu} -\frac{1}{2} g_{\mu \nu} R$ is the
 Einstein tensor, $\nabla$ represents the covariant derivative and 
$T_{\mu \nu}=\frac{2}{\pi}(g^{\gamma \sigma}F_{\mu \gamma}
F_{\nu \sigma}-\frac{1}{4}g_{\mu \nu}F_{\gamma \sigma}
F^{\gamma \sigma})$ is the Maxwell energy-momentum tensor. 

\section{GENERAL STATIC SOLUTION. ANALYSIS OF ITS
STRUCTURE}

\subsection{Field equations}

We want to consider now a spacetime which is both static and 
rotationally symmetric, implying the existence of a
timelike Killing vector $\partial/\partial t$ and a spacelike 
Killing vector $\partial/\partial\varphi$.
We will start working with the following ansatz for the metric 
\begin{equation}
  ds^2 = -\alpha^2 r^2 dt^2 + e^{-2\mu(r)}dr^2 
         +\frac{e^{2\mu(r)}}{\alpha^2} d\varphi^2 \:,
                               \label{MET_SYM}
\end{equation}
where the parameter $\alpha^2$ is, as we shall see, an appropriate  
constant proportional to the cosmological constant $\Lambda$.  
It is introduced in order to have metric components with  
dimensionless units and an asymptotically anti-de Sitter  
spacetime.
The motivation for this choice for the metric gauge  
[$g_{tt} \propto- r^2$ and 
$(g_{rr})^{-1} \propto g_{\varphi \varphi}$]  
instead of the usual ``Schwarzschild'' gauge [$(g_{rr})^{-1} =-g_{tt}$  
and  $g_{\varphi \varphi}=r^2$] comes from the fact that we are 
looking for magnetic solutions. Indeed, let us first remember  
that the Schwarzschild gauge is usually an appropriate choice  
when we are interested on electric solutions.  
Now, we focus on the well known fact that the electric field  
is associated with the time component, $A_t$, of the vector  
potential while the magnetic field is associated with 
the angular component $A_{\varphi}$. From 
the above facts, one can expect that 
a magnetic solution can be written in a metric gauge in which 
the components $g_{tt}$ and $g_{\varphi \varphi}$ interchange 
their roles relatively to those present in the ``Schwarzschild'' 
 gauge used to describe electric solutions.   
This choice will reveal to be a good one to find solutions since  
the dilaton and graviton will decouple from each other 
on the fields equations (\ref{EQUACAO_MET}) and
(\ref{EQUACAO_DIL}).
 However, as we will see, for some values of the Brans-Dicke parameter 
$\omega$ it is not the good coordinate system to interpret the solutions.

We now assume that the only non-vanishing components of the 
vector potential are $A_t(r)$ and $A_{\varphi}(r)$, i.e., 
\begin{equation}
A=A_tdt+A_{\varphi}d{\varphi}\:.  
\label{Potential}
\end{equation}
This implies that the non-vanishing components of the anti-symmetric
Maxwell tensor are $F_{tr}$ and $F_{r \varphi}$. Use of
metric (\ref{MET_SYM}) and equation (\ref{EQUACAO_MET}) yields
the following set of equations
\begin{eqnarray} 
 & & \hskip -0.4cm  \phi_{,rr}
       + 2 \phi_{,r} \mu_{,r}
       - (\omega+2) (\phi_{,r})^2
       - \frac{\mu_{,rr}}{2}
       - (\mu_{,r})^2+ \frac{1}{4} \Lambda e^{-2\mu}
      \nonumber \\
  & &  \hskip 3.5cm
      =\frac{\pi}{2 \alpha^2} \frac{e^{-2\mu}}{r^2} T_{tt} \:,
                                    \label{MET_00}  \\
  & &  \hskip -0.4cm
       - \phi_{,r} \mu_{,r}
       - \frac{\phi_{,r}}{r}
       - \omega (\phi_{,r})^2
       + \frac{\mu_{,r}}{2r}
       - \frac{1}{4} \Lambda e^{-2\mu}=
                  \frac{\pi}{2} T_{rr} \:,
                                    \label{MET_11}  \\
  & &   \hskip -0.4cm
       \phi_{,rr}
      + \phi_{,r} \mu_{,r}
      + \frac{\phi_{,r}}{r}
      - (\omega+2) (\phi_{,r})^2
      - \frac{\mu_{,r}}{2r}
      + \frac{1}{4} \Lambda e^{-2\mu}
             \nonumber \\
  & &  \hskip 3.5cm
      =-\frac{\pi}{2} \alpha^2 e^{-4\mu} T_{\varphi \varphi} \:,
                                           \label{MET_22}  \\ 
  & &  \hskip -0.4cm 
         0=\frac{\pi}{2} T_{t \varphi}=
                           e^{2\mu}F_{tr}F_{\varphi r} \:,
                                           \label{MET_02} 
\end{eqnarray}
where ${}_{,r}$ denotes a derivative with respect to $r$.
In addition, replacing the metric (\ref{MET_SYM}) into equations 
(\ref{EQUACAO_MAX}) and (\ref{EQUACAO_DIL}) yields 
\begin{eqnarray}
 & &  \partial_r {\bigl [} e^{-2 \phi}r(F^{t r}
      +F^{\varphi r}) {\bigr ]}=0\:,          
                                 \label{MAX_0} \\
& &  \omega \phi_{,rr}
      + 2 \omega \phi_{,r} \mu_{,r}
      + \omega \frac{\phi_{,r}}{r}
      - \omega (\phi_{,r})^2
      + \frac{\mu_{,r}}{r}
      + \frac{\mu_{,rr}}{2}
       \nonumber \\
  & &  \hskip 1cm
      + (\mu_{,r})^2
      -\frac{1}{4} \Lambda e^{-2\mu}=
       \frac{1}{4}e^{-2\mu}F^{\gamma \sigma}F_{\gamma \sigma} \:.
                                    \label{EQ_DIL}  
\end{eqnarray}

\subsection{\label{sec:stat_sol}The general static solution. 
Causal structure}

Equations (\ref{MET_00})-(\ref{EQ_DIL}) are  valid for a static and
rotationally symmetric spacetime. One sees that equation
(\ref{MET_02}) implies that the electric and
magnetic fields cannot be simultaneously  non-zero, i.e., there 
is no static dyonic solution. In this work we will consider the 
magnetically charged case alone ($A_t=0,\,A_{\varphi} \neq 0$).
For  purely electrically charged solutions of the theory see
\cite{OscarLemos}.
So, assuming vanishing electric field, one has from Maxwell 
equation (\ref{MAX_0}) that 
\begin{equation}
F_{\varphi r}=\frac{\chi_{\rm m}}{4 \alpha^2 r}e^{2 \phi},
\label{MAX_1}
\end{equation}
where $\chi_{\rm m}$ is an integration constant which measures
the intensity of the magnetic field source. 
To proceed we shall first consider the case $\omega\neq -1$.
Adding equations (\ref{MET_11}) and (\ref{MET_22}) yields
$\phi_{,rr}=2(\omega+1)(\phi_{,r})^2$,
and so the dilaton field is given by
\begin{equation}
  e^{-2\phi}= (\alpha r)^{\frac{1}{\omega+1}}, 
                       \quad\quad \omega \neq -1 \:,
                                    \label{DILATAO}
\end{equation}
where $\alpha$ is a generic constant which will be appropriately 
defined below in  equation (\ref{COSMOL}).
 The 1-form vector potential 
$A=A_{\mu}(r)dx^{\mu}$  is then
\begin{equation}
A=-\frac{1}{4 \alpha^2}\chi_{\rm m}(\omega+1)
(\alpha r)^{-\frac{1}{\omega+1}} d \varphi \:,
                           \quad\quad \omega \neq -1\:.
                                    \label{VEC_POTENT}
\end{equation}
Replacing solutions (\ref{MAX_1})-(\ref{VEC_POTENT}) 
into equations (\ref{MET_00})-(\ref{EQ_DIL}) allows us to find
the $e^{2\mu(r)}$ function for 
$\omega\neq\{-2,-3/2,-1\}$; $\omega =-2$
and $\omega =-3/2$, respectively
\begin{eqnarray}
e^{2\mu(r)} &=&
            (\alpha r)^2 + 
           \frac{b}{(\alpha r)^{\frac{1}{\omega+1}}}
           -\frac{k\chi_{\rm m}^2} 
            {(\alpha r)^{\frac{2}{\omega+1}}}\:,
                                 \label{MET_TODOS}  \\
 e^{2\mu(r)} &=&  {\biggl (}1-\frac{\chi_{\rm m}^2}{4}
           \ln{r}{\biggl )}r^2 -br \:, 
                                         \label{MET-2}  \\
 e^{2\mu(r)} &=&  r^2[-\Lambda \ln (br)-\chi_{\rm m}^2 r^2] \:,
                                         \label{MET-3/2}
\end{eqnarray}
where $b$ is a constant of integration related with the mass
of the solutions, as will be shown, and
$k=\frac{(\omega+1)^2}{8\alpha^2(\omega+2)}$.
For $\omega \neq -2,-3/2,-1 \:$ $\alpha$ is defined as
\begin{equation}
 \alpha =\sqrt{ {\biggl |} \frac{(\omega+1)\Lambda}
     {(\omega+2)(2\omega+3)} {\biggr |}}.
                                 \label{COSMOL}
\end{equation}
For $\omega=-2,-3/2$ we set $\alpha=1$.
For $\omega=-2$ equations (\ref{MET_00}) and (\ref{MET_11})
imply $\Lambda=-\chi_{\rm m}^2/8$.

Now, we consider the case $\omega=-1$. From equations 
(\ref{MET_00})-(\ref{EQ_DIL})
 it follows that $\mu=C_1$,
$\phi=C_2$, where $C_1$ and $C_2$ are constants of integration, 
and that the cosmological constant and magnetic source are both 
zero, $\Lambda=0=\chi_{\rm m}$. So, for $\omega=-1$ the metric 
gives simply the 3D Minkowski spacetime and the 
dilaton is constant,
as occurred in the uncharged case 
\cite{Sa_Lemos_Static,Sa_Lemos_Rotat}. 

Now, we must analyse carefully the radial dependence of the
 $e^{2\mu(r)}$ function defined in equations 
(\ref{MET_TODOS})-(\ref{MET-3/2}) which, recall, is related to 
the metric components
through the relations $g_{rr}=e^{-2\mu}$ and 
$g_{\varphi \varphi}=e^{2\mu}/\alpha^2$.
The shape of the $e^{2\mu(r)}$ function depends on the values of the
Brans-Dicke parameter $\omega$ and on the values of the parameter
$b$ (where $b$ is related to the mass of the solution as we shall
see in section \ref{sec:MJQ}). Nevertheless, we can group  the 
values of $\omega$ 
and $b$ into a small number of cases for which the $e^{2\mu(r)}$ function
 has the same behavior. The general shape of the $e^{2\mu(r)}$ function
for these cases is drawn in the Appendix. 
Generally, the $e^{2\mu(r)}$ function
can take positive or negative values depending on the value of the 
coordinate $r$. However, when $e^{2\mu(r)}$ is negative,
the metric components $g_{rr}=e^{-2\mu}$ and 
$g_{\varphi \varphi}=e^{2\mu}/\alpha^2$ become
simultaneously negative  and this 
leads to an apparent change of signature from $+1$ to $-3$. This 
strongly indicates that we are using an incorrect extension 
and that we should choose a different continuation to describe  
the region where the change of signature occurs \cite{HW,Horne_Horow}.
Moreover, analysing the null and timelike geodesic motion we conclude 
that null and timelike particles can never pass through the
 zero of $e^{2\mu(r)}$, $r_+$ (say), from the region where $g_{rr}$ is 
positive into the region where $g_{rr}$ is negative.
This suggests that one can introduce a new coordinate  
system in order to obtain a  spacetime which is geodesically complete for 
the region where both $g_{rr}$ and $g_{\varphi \varphi}$
are positive \cite{HW,Horne_Horow}. 
That is our next step. Then, in section \ref{sec:geod}, we will check the 
completeness of the spacetimes.
The Brans-Dicke theories can be classified into seven different cases
that we display and study below. 

\subsubsection*{{\bf (i)} Brans-Dicke theories with $-1<\omega<+\infty$}

The shape of the $e^{2\mu(r)}$ function 
is drawn in Fig. 1(a). We see that for $0<r<r_+$ 
(where $r_+$ is the zero of $e^{2\mu(r)}$)
$g_{rr}$ and $g_{\varphi \varphi}$ become
simultaneously negative  and this 
leads to an apparent change of signature.
One can however introduce a new radial coordinate $\rho$ 
so that the  spacetime is geodesically complete for 
the region where both $g_{rr}$ and $g_{\varphi \varphi}$
are positive,
\begin{eqnarray}
\rho^2=r^2-r_+^2 \:. 
                                        \label{Transf_1}
\end{eqnarray}
With this coordinate transformation, the
spacetime generated by the static magnetic point source
is finally  given by
\begin{eqnarray}
 ds^2 = -\alpha^2 r^2(\rho) dt^2
     + \frac{\rho^2}{r^2(\rho)}\frac{1}{f(\rho)} d \rho^2
        + \frac{f(\rho)}{\alpha^2}
            d\varphi^2 \:,
                     \label{Met_1}  
\end{eqnarray}
where $0\leq\rho<\infty$, and function $f(\rho)$, which is 
always positive except at 
$\rho=0$ where it is zero, is given by
\begin{eqnarray} 
f(\rho)=  \alpha^2 r^2(\rho) + 
      \frac{b}{[\alpha^2 r^2(\rho)]^{\frac{1}{2(\omega+1)}}}
           -\frac{k\chi_{\rm m}^2}
    {[\alpha^2 r^2(\rho)]^{\frac{1}{\omega+1}}} \:.  
                            \label{f}   
\end{eqnarray} 
Along this section \ref{sec:stat_sol}, in cases {\bf(i)} and  {\bf(iii)}
we will use the definition $r^2(\rho) \equiv \rho^2 + r_+^2$ 
in order to shorten the formulas. 
This spacetime has no horizons and so there are no magnetic 
black hole solutions, only magnetic point sources. 
In three dimensions, the presence of a 
curvature singularity is revealed by the scalar 
$R_{\mu\nu}R^{\mu\nu}$
\begin{eqnarray}
 \hspace{-0.2cm} & & \hspace{-0.2cm} 
        R_{\mu\nu}R^{\mu\nu} = 
                                 \nonumber \\  
\hspace{-0.2cm} & & \hspace{-0.2cm} 
        - \frac{4 \omega}{(\omega+1)^2}
       \frac{b \alpha^4}
       {[\alpha^2 r^2(\rho)]^{\frac{2\omega+3}{2(\omega+1)}}} 
    + \frac{(2\omega^2+4\omega+3)b^2\alpha^4}{2(\omega+1)^4
       [\alpha^2 r^2(\rho)]^{\frac{2\omega+3}{\omega+1}}}  
                                                \nonumber \\         
\hspace{-0.2cm} & &  \hspace{-0.2cm}    
       +\frac{(\omega-1)}{(\omega+1)^2}
       \frac{8 k \chi_{\rm m}^2 \alpha^4}      
       {[\alpha^2 r^2(\rho)]^{\frac{\omega+2}{\omega+1}}}   
      -\frac{(\omega^2+2\omega+2)}{(\omega+1)^4}
       \frac{k \chi_{\rm m}^2 b \alpha^4}      
       {[\alpha^2 r^2(\rho)]^{\frac{4\omega+7}{2(\omega+1)}}}  
                                                  \nonumber \\
\hspace{-0.2cm} & &  \hspace{-0.2cm}              
        -\frac{(\omega^2+2\omega+3)}{(\omega+1)^4}  
             \frac{k^2 \chi_{\rm m}^4 \alpha^4}      
      {[\alpha^2 r^2(\rho)]^{\frac{2(\omega+2)}{\omega+1}}}
       + 12\alpha^4   \:.               
                             \label{R-2-3/2-1}
\end{eqnarray} 
This scalar does not diverge for any value
of $\rho$ (if $\omega>-1$). Therefore, spacetime 
(\ref{Met_1}) has no curvature singularities.
However, it has a conic geometry with a 
conical singularity at $\rho=0$. In fact, in the vicinity of 
$\rho=0$,  metric (\ref{Met_1}) is written as 
\begin{eqnarray}
 ds^2 \sim -\alpha^2  r_+^2 dt^2 + \frac{\nu}{\alpha r_+}
       d\rho^2+(\alpha r_+ \nu)^{-1} \rho^2  d\varphi^2 \:,
                                 \label{Met_1_0}  
\end{eqnarray}
with $\nu$ given by  
\begin{eqnarray}
\nu=  {\biggl [} 
        \alpha r_+ - \frac{b 
         (\alpha r_+)^{-\frac{\omega+2}{\omega+1}}}
           {2(\omega+1)} 
       +\frac{k\chi_{\rm m}^2}{\omega+1} 
      (\alpha r_+)^{-\frac{\omega+3}{\omega+1}} 
       {\biggr ]}^{-1}\:.
                                        \label{T_Con_1}
\end{eqnarray}
So, there is indeed a conical singularity at $\rho=0$ 
since as the radius $\rho$ 
tends to zero, the limit of the 
ratio ``circumference/radius'' is not $2\pi$. 
The period of  coordinate $\varphi$ associated with this 
conical singularity is
\begin{eqnarray}
{\rm Period}_{\varphi}=
2 \pi {\biggl [}  \lim_{\rho \rightarrow 0} 
\frac{1}{\rho} \sqrt{\frac{g_{\varphi \varphi}} 
{g_{\rho \rho}}} {\biggr ]}^{-1}
=2 \pi \nu \:.
                                        \label{T_Con}
\end{eqnarray}
From (\ref{Met_1_0})-(\ref{T_Con}) one concludes that  
in the vicinity of the origin, metric (\ref{Met_1}) describes  
a spacetime which is locally flat but has a conical 
singularity with an angle deficit  
$\delta \varphi=2\pi(1-\nu)$. 

Before closing this case,
one should mention the particular $\omega=0$ case 
of Brans-Dicke theory since this theory is related 
(through dimensional reduction) 
to 4D General Relativity with one Killing vector 
studied in \cite{OscarLemos_string}.

\subsubsection*{{\bf (ii)} Brans-Dicke theory with $\omega=\pm \infty$}

The Brans-Dicke theory defined by $\omega=\pm \infty$ reduces
to the spacetime generated by a static 
magnetic point source in 3D Einstein-Maxwell 
theory with $\Lambda<0$ studied in detail in 
\cite{CL1}, \cite{HW}-\cite{OscarLemos_BTZ}.
The behavior of this spacetime is quite similar to those
described in case {\bf (i)}. It has a conical singularity at the origin
and no horizons.

\subsubsection*{{\bf (iii)} Brans-Dicke theories with $-\infty<\omega<-2$} 

For this range of the Brans-Dicke parameter we have to consider
separately the case (1) $b>0$ and (2) $b<0$, where $b$ is the mass parameter.

(1) If $b>0$ the shape of the $e^{2\mu(r)}$ function 
is drawn in Fig. 1(b). Both $g_{rr}$ and 
$g_{\varphi \varphi}$ are always positive (except at $r=0$) and 
there is no apparent change of signature. Hence, for this range of
parameters, the
spacetime is correctly described by equations (\ref{MET_SYM}) and
(\ref{MET_TODOS}).
There are no horizons and so no magnetic black holes,  but at 
$r=0$ the scalar 
$R_{\mu\nu}R^{\mu\nu}$ diverges (in equation (\ref{R-2-3/2-1}) put
$r_+=0$ and replace $\rho$ by $r$). Therefore, at $r=0$ one has the 
presence of a naked curvature singularity.

(2) If $b<0$ the shape of the $e^{2\mu(r)}$ function defined in 
(\ref{MET_TODOS}) is sketched in Fig. 1(c).
There occurs an apparent change of signature for $0<r<r_+$. 
Proceeding exactly as we did in case {\bf (i)} 
one can however introduce the new radial coordinate $\rho$
defined in  (\ref{Transf_1}) and obtain the geodesically complete 
spacetime described by (\ref{Met_1}) and (\ref{f}) 
(where now  $\omega<-2$). This spacetime has no horizons
and the scalar $R_{\mu\nu}R^{\mu\nu}$ given by 
(\ref{R-2-3/2-1}) does not diverge for any value
of $\rho$ and so  no curvature singularities are present.
The spacetime has a conical 
singularity at $\rho=0$ corresponding to an angle deficit   
$\delta \varphi=2\pi(1-\nu)$, where $\nu$ is defined in  
(\ref{T_Con_1}).

\subsubsection*{{\bf (iv)} Brans-Dicke theory with $\omega=-2$} 

The shape of the $e^{2\mu(r)}$ function is drawn in Fig. 2(a).
There is an apparent change of signature for $r>r_+$,
where $r_+$ is the zero of $e^{2\mu(r)}$.
We can however introduce a new coordinate system that 
will allow us to conclude that the spacetime is complete for 
$0 \leq r \leq r_+$. First, we introduce the radial coordinate 
$R=1/r$. With this new coordinate $[g_{RR}(R)]^{-1}$ has a shape similar
to the one shown in Fig. 1(a). Finally, we set a second 
coordinate transformation given by
$\rho^2=R^2-R_+^2$, where $R_+=1/r_+$ is the zero of $(g_{RR})^{-1}$.
Use of these coordinate transformations, together with 
equations (\ref{MET_SYM}) and (\ref{MET-2}),
allows us to  write the
spacetime generated by the static magnetic point source as
\begin{eqnarray}
 ds^2  = 
      -\frac{\alpha^2} {R^2(\rho)} dt^2
     + \frac{\rho^2}{[R^2(\rho)]^3} \frac{1}{h(\rho)} d\rho^2
          + \frac{h(\rho)}{\alpha^2}
            d\varphi^2 \:, 
                                 \label{Met_2}  
\end{eqnarray}
where $0\leq\rho<\infty$ and the function $h(\rho)$ is given by
\begin{eqnarray} 
h(\rho) = { {\biggl (} 1+\frac{\chi_{\rm m}^2}{8}
      \ln{[R^2(\rho)]} {\biggr )} R^{-2}(\rho) 
            -b R^{-1}(\rho)}  \:. 
                            \label{g_2}
\end{eqnarray} 
This function $h(\rho)$  is 
always positive except at $\rho=0$ where it is zero.
Hence, the spacetime described by equation (\ref{Met_2}) and (\ref{g_2}) 
has no horizon. 
Along this section \ref{sec:stat_sol}, in cases {\bf(iv)}-{\bf(vii)}
we will use the definition $R^2(\rho) \equiv \rho^2 +R_+^2$ 
in order to shorten the formulas. 

The scalar 
$R_{\mu\nu}R^{\mu\nu}$ is given by
\begin{eqnarray}
R_{\mu\nu}R^{\mu\nu} \hspace{-0.1cm} &=&  \hspace{-0.1cm}  
       \chi_{\rm m}^4 {\biggl [}\frac{3}{4}\ln^2[R(\rho)]        
         + \frac{5}{4} \ln [R(\rho)] +9{\biggr ]}+
                                             \nonumber \\    
\hspace{-0.1cm} & &  \hspace{-0.1cm} 
-\chi_{\rm m}^2 {\biggl [}6 \ln [R(\rho)]  + 
   \frac{4 \ln [R(\rho)]}{R(\rho)} +\frac{3}{R(\rho)}+5{\biggr ]}+
                                               \nonumber \\    
\hspace{-0.1cm} & &  \hspace{-0.1cm} 
 + 8+\frac{32}{R(\rho)}+\frac{6}{R^2(\rho)}        
   \:.
                                \label{R-2} 
\end{eqnarray}
This scalar diverges for $\rho=+\infty$ and so there is a curvature 
singularity at $\rho=+\infty$. 
Besides, the  spacetime described by 
(\ref{Met_2}) and (\ref{g_2}) has a 
conical singularity at $\rho=0$ with coordinate $\varphi$ 
having a period defined in
equation (\ref{T_Con}), 
\begin{eqnarray}
{\rm Period}_{\varphi}=2 \pi {\biggl [}r_+ {\biggl (}
           \frac{\chi_{\rm m}^2}{8}-1+\frac{b}{2r_+}
       -\frac{\chi_{\rm m}^2}{4} \ln{r_+}
       {\biggr )}  {\biggr ]}^{-1}\:.
                                        \label{T_Con_3}
\end{eqnarray}
So, near the origin,  metric (\ref{Met_2}) and (\ref{g_2}) 
describe a spacetime which is locally flat but has a conical 
singularity at $\rho=0$ with an angle deficit  
$\delta \varphi=2\pi-{\rm Period}_{\varphi}$.

\subsubsection*{{\bf (v)}  Brans-Dicke theories with $-2<\omega<-3/2$}  

For this range of the Brans-Dicke parameter we have again  to consider
separately the case (1) $b>0$ and (2) $b<0$, 
where $b$ is the mass parameter.

(1) If $b>0$ the shape of the $e^{2\mu(r)}$ function defined in 
(\ref{MET_TODOS})  is similar to the one of 
case {\bf (iv)} and  sketched in Fig. 2(a).
So, proceeding as in case  {\bf (iv)}, we find that the
spacetime generated by the static magnetic point source is
given by (\ref{Met_2}) with function $h(\rho)$ defined by
\begin{eqnarray} 
h(\rho)  \hspace{-0.1cm}=  \hspace{-0.1cm} 
     \frac{\alpha^2} {R^2(\rho)} + 
                              \hspace{-0.1cm}
      b {\biggl (} \frac{\alpha^2} {R^2(\rho)} 
      {\biggr )}^{-\frac{1}{2(\omega+1)}}
   \hspace{-0.4cm}
        -k\chi_{\rm m}^2  {\biggl (}
    \frac{\alpha^2} {R^2(\rho)} 
     {\biggr )}^{-\frac{1}{\omega+1}}
      \hspace{-0.2cm} ,  
                            \label{g}   
\end{eqnarray} 
which is always positive except at $\rho=0$ where it is zero.
Hence, the spacetime described by equations (\ref{Met_2}) and 
(\ref{g}) has no horizons.

The scalar $R_{\mu\nu}R^{\mu\nu}$ is given by (\ref{R-2-3/2-1})
as long as we replace function $r^2(\rho)$ by 
$R^{-2}(\rho) \equiv (\rho^2 +R_+^2)^{-1}$.
There is a curvature 
singularity at $\rho=+\infty$.

Near the origin, equations (\ref{Met_2}) and (\ref{g}) 
describe a spacetime which is locally flat but has a conical 
singularity at $\rho=0$ with an angle deficit  
$\delta \varphi=2\pi-{\rm Period}_{\varphi}$, 
with  ${\rm Period}_{\varphi}$ defined in
equation (\ref{T_Con}),
\begin{eqnarray}
\hspace{-0.8cm} & & \hspace{-0.8cm}
 {\rm Period}_{\varphi} =       \nonumber \\
   \hspace{-0.1cm} & & \hspace{-0.1cm}              
2 \pi   {\biggl [} 
        \alpha r_+ - \frac{b 
         (\alpha r_+)^{-\frac{\omega+2}{\omega+1}}}
           {2(\omega+1)} 
       +\frac{k\chi_{\rm m}^2}{\omega+1} 
      (\alpha r_+)^{-\frac{\omega+3}{\omega+1}} 
       {\biggr ]}^{-1} \hspace{-0.3cm}.                
                        \label{T_Con_2}
\end{eqnarray}

(2) If $b<0$ the $e^{2\mu(r)}$ function can have a shape 
similar to the one sketched in Fig. 2(b) or similar to Fig. 2(c), 
depending on the values of the range.
We will not proceed further with the study of this case 
since it has a rather exotic spacetime structure.

\subsubsection*{{\bf (vi)}  Brans-Dicke theory with $\omega=-3/2$}  

The shape of the $e^{2\mu(r)}$ function defined in 
(\ref{MET-3/2})  is similar to the one of 
case  {\bf (iv)} and  sketched in Fig. 2(a).
So, proceeding as in case  {\bf (iv)}, we conclude that the
spacetime generated by the static magnetic point source is
given by (\ref{Met_2}) with function $h(\rho)$ defined by
\begin{eqnarray} 
\hspace{-0.2cm}  h(\rho) =  R^{-2}(\rho) {\biggl (} 
      \frac{\Lambda}{2} \ln [b^{-2}R^2(\rho)]
      -\chi_{\rm m}^2 R^{-2}(\rho)
           {\biggr )} \:, 
                             \label{g_3}
\end{eqnarray} 
which is always positive except at $\rho=0$ where it is zero.
Hence, the spacetime described by equations (\ref{Met_2}) and 
(\ref{g_3}) has no horizons. 

The scalar 
$R_{\mu\nu}R^{\mu\nu}$ is
\begin{eqnarray}
\hspace{-0.8cm} & &  \hspace{-0.8cm}  R_{\mu\nu}R^{\mu\nu} =
       \Lambda^2 {\bigl [}12\ln^2[bR(\rho)]+20 \ln[bR(\rho)]+9{\bigr ]}+
                                         \nonumber \\
  \hspace{-0.3cm} & & \hspace{-0.3cm} 
           - \Lambda \chi_{\rm m}^2 R^2(\rho) {\biggr [} 
     5 \ln[bR(\rho)]+\frac{9}{2}
   {\biggr ]}+\frac{9}{16}\chi_{\rm m}^4 R^4(\rho) \:.  
                                   \label{R-3/2}
\end{eqnarray}
There is a curvature 
singularity at $\rho=+\infty$.

Near the origin, equations (\ref{Met_2}) and (\ref{g_3}) 
describe a spacetime which is locally flat but has a conical 
singularity at $\rho=0$ with an angle deficit  
$\delta \varphi=2\pi-{\rm Period}_{\varphi}$, 
with ${\rm Period}_{\varphi}$ defined in
equation (\ref{T_Con}),
\begin{eqnarray}
{\rm Period}_{\varphi}=2 \pi {\biggl [}r_+ {\biggl (}
            \frac{\Lambda}{2} +          
            \chi_{\rm m}^2 r_+^2 {\biggr )}  {\biggr ]}^{-1}  \:.
                                        \label{T_Con_4}
\end{eqnarray}

\subsubsection*{{\bf (vii)} Brans-Dicke theories with $-3/2<\omega<-1$}  

The shape of the $e^{2\mu(r)}$ function is sketched in Fig. 2(a)
and is similar to the one of case  {\bf (iv)}.
So, proceeding as in case  {\bf (iv)}, we find that the
spacetime generated by the static magnetic point source is
given by (\ref{Met_2}) with function $h(\rho)$ defined by
(\ref{g}). There are no horizons and there is no curvature singularity
[the scalar $R_{\mu\nu}R^{\mu\nu}$ is given by (\ref{R-2-3/2-1})
if we replace function $r^2(\rho)$ by 
$R^{-2}(\rho)$].
The spacetime has a conical singularity at $\rho=0$ corresponding 
to an angle deficit $\delta \varphi=2\pi-{\rm Period}_{\varphi}$,  
 where ${\rm Period}_{\varphi}$ is defined in  
(\ref{T_Con_2}).

\subsection{\label{sec:geod}Geodesic structure}

We want to confirm that the spacetimes described by (\ref{MET_SYM}),
 (\ref{Met_1}) and  (\ref{Met_2})
are both null and timelike geodesically complete.
The equations governing the geodesics can be derived from the
 Lagrangian
\begin{equation}
{\cal{L}}=\frac{1}{2}g_{\mu\nu}\frac{dx^{\mu}}{d \tau}
       \frac{dx^{\nu}}{d \tau}=-\frac{\delta}{2}\:,
                                 \label{LAG)}  \\
\end{equation}
where $\tau$ is an affine parameter along the geodesic which, 
for a timelike geodesic, can be identified with the proper 
time of the particle along the geodesic. For a null geodesic 
one has $\delta=0$ and for a timelike geodesic $\delta=+1$. 
From the Euler-Lagrange equations one gets that the generalized 
momentums associated with the time coordinate and angular 
coordinate are constants: $p_t=E$ and $p_{\varphi}=L$. The 
constant $E$ is related to the timelike Killing vector 
$(\partial/\partial t)^{\mu}$ which reflects the time 
translation invariance of the metric, while the constant 
$L$ is associated to the spacelike Killing vector 
$(\partial/\partial \varphi)^{\mu}$ which reflects the 
invariance of the metric under rotation. Note that since the 
spacetime is not asymptotically flat, the constants $E$ and 
$L$ cannot be interpreted as the  energy and angular 
momentum at infinity. 

From the metric we can derive the radial geodesic, 
\begin{eqnarray}
\dot{\rho}^2=-\frac{1}{g_{\rho\rho}}
      \frac{E^2 g_{\varphi \varphi}+L^2 g_{tt}}
              { g_{tt} g_{\varphi \varphi} } 
       -\frac{\delta}{g_{\rho\rho}} \:. 
                                        \label{GEOD_1}
\end{eqnarray}
Next, we analyse this geodesic equation for each
of the seven  cases defined in the last section. 
Cases {\bf (i)}, {\bf (ii)} and {\bf (iii)} have identical
geodesic structure, and cases {\bf (iv)}-{\bf (vii)} also.

\subsubsection*{{\bf (i)} Brans-Dicke theories with $-1<\omega<+\infty$}

Using the two useful relations 
$g_{tt} g_{\varphi \varphi}=-\rho^2/g_{\rho\rho}$ and 
$g_{\varphi \varphi}=[\rho^2 / (\rho^2+r_+^2)](\alpha^2 g_{\rho\rho})^{-1}$,
 we can write equation (\ref{GEOD_1}) as  
\begin{eqnarray}
\rho^2 \dot{\rho}^2= {\biggl [}
\frac{E^2}{\alpha^2}\frac{1}{\rho^2 + r_+^2}
-\delta {\biggr ]} \frac{\rho^2}{g_{\rho \rho}}     
            +L^2 g_{tt} \:. 
                                        \label{Geod_1}
\end{eqnarray}
Noticing that $1/g_{\rho\rho}$ is always positive for $\rho>0$ and 
zero for $\rho=0$, and that $g_{tt}<0$ we conclude  
the following about the null geodesic motion ($\delta=0$).
The first term in (\ref{Geod_1}) is positive
(except at $\rho=0$ where it vanishes),  while the 
second term is always negative. We can then conclude that 
spiraling ($L \neq 0$) null particles coming in from an arbitrary 
point are scattered at the turning point $\rho_{\rm tp} > 0$ and 
spiral back to infinity. If the angular momentum L of the null 
particle is zero it hits the origin (where there is a conical
singularity) with vanishing velocity.

Now we analyze the  timelike geodesics ($\delta=+1$). 
Timelike geodesic motion is possible only if the energy of the 
particle satisfies $E > \alpha r_+$. In this case,
spiraling timelike particles are bounded between two turning 
points that satisfy $\rho_{\rm tp}^{\rm a} > 0$ and 
$\rho_{\rm tp}^{\rm b} < \sqrt{E^2/\alpha^2 - r_+^2}$, with
$\rho_{\rm tp}^{\rm b} \geq \rho_{\rm tp}^{\rm a}$. 
When the timelike particle has no angular momentum ($L=0$) 
there is a turning point located exactly at 
$\rho_{\rm tp}^{\rm b}=\sqrt{E^2/\alpha^2 - r_+^2}$
and it hits the origin $\rho=0$.
Hence, we confirm that the spacetime described by  
equation (\ref{Met_1}) is both timelike and null geodesically 
complete.

\subsubsection*{{\bf (ii)} Brans-Dicke theory with $\omega=\pm \infty$}

The geodesic structure of the spacetime generated by a static 
magnetic point source in 3D Einstein-Maxwell 
theory with $\Lambda<0$ has been studied in detail in 
\cite{OscarLemos_BTZ}.
The behavior of this geodesic structure is quite similar to the one
described in case {\bf (i)}.
In particular, the spacetime is both timelike and null geodesically 
complete. 
 
\subsubsection*{{\bf (iii)} Brans-Dicke theories with $-\infty<\omega<-2$}

For this range of the Brans-Dicke parameter we have to consider
separately the case (1) $b>0$ and (2) $b<0$, where $b$ is the mass parameter.

(1) If $b>0$ the motion of null and timelike geodesics is 
correctly described by equation (\ref{Geod_1}) if we replace 
$\rho$ by $r$ and put $r_+ =0$.
Hence, the null and timelike geodesic  motion has the same 
feature as the one described in the above case {\bf (i)}. 
In the above statements we just 
have to replace $\rho$ by $r$, put $r_+ =0$ and remember that
at the origin there is now a naked curvature singularity rather than
a conical singularity.

 (2) If $b<0$ the motion of null and timelike geodesics is exactly 
described by (\ref{Geod_1}) and the statements presented in case {\bf (i)}
apply directly to this case.

\subsubsection*{{\bf (iv)}, {\bf (v)}, {\bf (vi)}, {\bf (vii)} 
Brans-Dicke theories with $-2 \leq \omega < -1$ }

In order to study the geodesic motion of spacetime described by
equations (\ref{Met_2}), one first notices that 
$g_{tt} g_{\varphi \varphi}=-[g_{\rho\rho}(\rho^2+R_+^2)^2 / \rho^2]^{-1}$ 
and $g_{\varphi \varphi}=[\rho^2 / (\rho^2+R_+^2)^3]
(\alpha^2 g_{\rho\rho})^{-1}$. Hence
 we can write equation (\ref{GEOD_1}) as 
\begin{eqnarray}
\rho^2 \dot{\rho}^2= {\biggl [}
\frac{E^2}{\alpha^2}\frac{1}{\rho^2 + R_+^2}
-\delta {\biggr ]} \frac{\rho^2}{g_{\rho \rho}}     
            + (\rho^2 + R_+^2)^2 L^2 g_{tt} \:. 
                                        \label{Geod_2}
\end{eqnarray}
One concludes that the geodesic motion of null and timelike particles 
has the same feature as the one
described in case {\bf (i)} after (\ref{Geod_1}) if in the statements
we replace $r_+$ by $R_+$. The only
difference is that on $\rho=+\infty$ there is now a curvature
singularity for cases {\bf (iv)} $\omega=-2$, 
{\bf (v)} $-2<\omega<-3/2$ and {\bf (vi)} $\omega=-3/2$.
In case {\bf (v)} $-2<\omega<-3/2$, if $b<0$ (as we saw in last section) 
the spacetime has an exotic structure and so we do not study it.

So, we confirm  that the spacetimes described by  
equations (\ref{Met_2}) are also  both timelike and 
null geodesically complete.

\section{THE GENERAL ROTATING SOLUTION}

\subsection{The generating technique}

Now, we want to endow our spacetime solution with a global
rotation, i.e., we want to add angular momentum to the spacetime.
In order to do so we perform
the following rotation boost in the $t$-$\varphi$ plane 
(see e.g. \cite{Sa_Lemos_Rotat}-\cite{Zanchin_Lemos}, 
\cite{HorWel})
\begin{eqnarray}
 t &\mapsto& \gamma t-\frac{\varpi}{\alpha^2} \varphi \:,
                                       \nonumber  \\
 \varphi &\mapsto& \gamma \varphi-\varpi t \:,
                                       \label{TRANSF_J}
\end{eqnarray}
where $\gamma$ and $\varpi$ are constant parameters.

\subsubsection*{{\bf (i)} Brans-Dicke theories with $-1<\omega<+\infty$}

Use of equation (\ref{TRANSF_J}) and
(\ref{Met_1}) gives the gravitational field 
generated by the rotating magnetic source 
\begin{eqnarray}
 \hspace{-1cm}  & & \hspace{-1cm}
 ds^2 = -\alpha^2 (\rho^2 + r_+^2) 
       (\gamma dt-\frac{\varpi}{\alpha^2} d\varphi)^2 + 
                                            \nonumber \\ 
   \hspace{-0.4cm}  & & \hspace{-0.4cm}
     + \frac{\rho^2}{(\rho^2 + r_+^2)}\frac{1}{f(\rho)} d \rho^2
     + \frac{f(\rho)}{\alpha^2} (\gamma d\varphi-\varpi dt)^2,
                     \label{Met_1_J}  
\end{eqnarray}
where $f(\rho)$ is defined in (\ref{f}).

The 1-form electromagnetic vector potential, $A=A_{\mu}(\rho)dx^{\mu}$, 
of the rotating solution is 
\begin{equation}
A=-\varpi A(\rho)dt +\gamma  A(\rho) d\varphi\:,
                                    \label{VEC_POTENT_J}
\end{equation}
where 
\begin{equation}
A(\rho)=-\frac{1}{4\alpha^2}\chi_{\rm m}(\omega+1)
[\alpha^2 (\rho^2+r_+^2)]^{-\frac{1}{2(\omega+1)}}\:.
                            \label{A_J}
\end{equation}

\subsubsection*{{\bf (ii)} Brans-Dicke theory with $\omega=\pm \infty$}

The spacetime generated by a rotating 
magnetic point source in 3D Einstein-Maxwell 
theory with $\Lambda<0$ has been obtained and studied in detail in 
\cite{OscarLemos_BTZ}.
 
\subsubsection*{{\bf (iii)} Brans-Dicke theories with $\omega<-2$}

Proceeding exactly as in case {\bf (i)}, we conclude that
the gravitational and electromagnetic fields 
generated by the rotating magnetic source are also
described by equations (\ref{Met_1_J})-(\ref{A_J}).

If $b>0$ we have to set $r_+ =0$ in equations (\ref{Met_1_J}) and 
(\ref{A_J}). 

\subsubsection*{{\bf (iv)} Brans-Dicke theory with $-\infty<\omega<-2$}

Use of equations (\ref{TRANSF_J}) and
(\ref{Met_2}) yields the gravitational field 
generated by the rotating magnetic source 
\begin{eqnarray}
 \hspace{-1cm}  & & \hspace{-1cm} 
      ds^2  =
      -\frac{\alpha^2} {(\rho^2 + R_+^2)}  
     (\gamma dt-\frac{\varpi}{\alpha^2} d\varphi)^2+ 
                                          \nonumber \\ 
 \hspace{-0.4cm}  & & \hspace{-0.4cm} 
     + \frac{\rho^2}{(\rho^2 + R_+^2)^3} \frac{1}{h(\rho)} d\rho^2
        + \frac{h(\rho)}{\alpha^2}
        (\gamma d\varphi-\varpi dt)^2, 
                                 \label{Met_2_J}  
\end{eqnarray}
where $h(\rho)$ is defined in (\ref{g_2}).

The 1-form vector potential is also given by 
(\ref{VEC_POTENT_J}) but now  one has 
\begin{equation}
A(\rho)=-\frac{1}{4 \alpha^2}\chi_{\rm m}(\omega+1)
[\alpha^{-2} (\rho^2 + R_+^2)]^{\frac{1}{2(\omega+1)}}\:.
                            \label{A_J_2}
\end{equation}

\subsubsection*{{\bf (v)}  Brans-Dicke theories with $-2<\omega<-3/2$}

If $b>0$, the gravitational and electromagnetic fields 
generated by the rotating magnetic source are
described by equations (\ref{Met_2_J}) and (\ref{A_J_2}),
with $h(\rho)$ defined in (\ref{g}).

\subsubsection*{{\bf (vi)}  Brans-Dicke theory with $\omega=-3/2$ }

The gravitational and electromagnetic fields 
generated by the rotating magnetic source are 
described by equations (\ref{Met_2_J}) and (\ref{A_J_2}),
with $h(\rho)$ defined in (\ref{g_3}).

\subsubsection*{{\bf (vii)} Brans-Dicke theories with $-3/2<\omega<-1$}
The gravitational and electromagnetic fields 
generated by the rotating magnetic source are 
described by equations (\ref{Met_2_J}) and (\ref{A_J_2}),
with $h(\rho)$ defined in (\ref{g}).

\vskip 3mm

In equations (\ref{Met_1_J}), (\ref{VEC_POTENT_J}) and (\ref{Met_2_J}) 
we choose $\gamma^2-\varpi^2 / \alpha^2=1$ because in this way  
when the angular momentum vanishes ($\varpi=0$) we have 
$\gamma=1$ and so we recover the static solution.

Solutions (\ref{Met_1_J})-(\ref{A_J_2}) represent
magnetically charged stationary spacetimes and also solve 
(\ref{ACCAO}). Analyzing the Einstein-Rosen bridge of the 
static solution one concludes that spacetime is not simply 
connected which implies that the first Betti number of the 
manifold is one, i.e., closed curves encircling the horizon 
cannot be shrunk to a point. So, transformations 
(\ref{TRANSF_J}) generate a new metric because they are not 
permitted global coordinate transformations \cite{Stachel}. 
Transformations (\ref{TRANSF_J}) can be done locally, but 
not globally. Therefore metrics 
(\ref{Met_1}), (\ref{Met_2}) and 
(\ref{Met_1_J})-(\ref{A_J_2}) can be locally mapped into 
each other but not globally, and such they are distinct.

\subsection{\label{sec:MJQ} Mass, angular momentum and 
charge of the solutions}

As we shall see the spacetime solutions describing 
the cases {\bf (i)} $-1<\omega<+\infty$, {\bf (ii)}  $\omega=\pm \infty$ and 
{\bf (iii)} $-\infty<\omega <-2$ are asymptotically anti-de Sitter.
This fact allows us to calculate the mass, angular momentum 
and the electric charge of the static and rotating solutions. 
To obtain these 
quantities  we apply the formalism  of Regge and Teitelboim 
\cite{Regge} 
(see also \cite{BTZ_Q},
\cite{Sa_Lemos_Static}-\cite{Lemos}).

\subsubsection*{{\bf (i)} Brans-Dicke theories with $-1<\omega<+\infty$}

We first write  (\ref{Met_1_J}) in the canonical form 
involving the lapse function $N^0(\rho)$ and the shift 
function $N^{\varphi}(\rho)$
\begin{equation}
     ds^2 = - (N^0)^2 dt^2
            + \frac{d\rho^2}{f^2}
            + H^2(d\varphi+N^{\varphi}dt)^2 \:,
                               \label{MET_CANON}
\end{equation}
where $f^{-2}=g_{\rho\rho}$, $H^2=g_{\varphi \varphi}$, 
 $H^2 N^{\varphi}=g_{t \varphi}$ and 
$(N^0)^2-H^2(N^{\varphi})^2=g_{tt}$.
Then, the action can be written in the Hamiltonian form as a 
function of the energy constraint ${\cal{H}}$, momentum constraint 
${\cal{H}}_{\varphi}$ and Gauss constraint $G$
\begin{eqnarray}
S \hspace{-0.2cm}  &=& \hspace{-0.2cm} 
     -\int dt d^2x[N^0 {\cal{H}}+N^{\varphi} 
         {\cal{H}_{\varphi}}
       + A_{t} G]+   {\cal{B}}         
                                     \nonumber \\
 \hspace{-0.2cm} &=& \hspace{-0.2cm}
         -\Delta t \int d\rho N \nu
        {\biggl [} \frac{2 \pi^2 e^{-2 \phi}}{H^3} 
        -4f^2(H \phi_{,\rho}e^{-2 \phi})_{,\rho}
                                      \nonumber \\
\hspace{-0.5cm}  & & \hspace{-0.5cm} 
        -2H \phi_{,\rho}(f^2)_{,\rho} e^{-2 \phi} 
        +2f(fH_{,\rho})_{,\rho}e^{-2 \phi}
                                      \nonumber \\
 \hspace{-0.5cm}  & & \hspace{-0.5cm} 
        +4 \omega H f^2 (\phi_{,\rho})^2e^{-2 \phi}
        -\Lambda H e^{-2 \phi}
       +\frac{2H}{f}e^{-2 \phi}(E^2+B^2){\biggr ]} 
                                        \nonumber \\
\hspace{-0.5cm}  & & \hspace{-0.5cm}
  + \Delta t \int d\rho N^{\varphi}\nu{\biggl [}
      {\bigl (}2 \pi e^{-2 \phi} {\bigr )}_{,\rho}
       +\frac{4H}{f}e^{-2 \phi}E^{\rho}B{\biggr ]}
                                            \nonumber \\
\hspace{-0.5cm}  & & \hspace{-0.5cm}
       + \Delta t \int d \rho A_t \nu {\biggl [}-\frac{4H}{f}
       e^{-2 \phi} \partial_{\rho} E^\rho{\biggr ]} +{\cal{B}} \:,
                               \label{ACCAO_CANON}
\end{eqnarray}
where $N=\frac{N^0}{f}$, 
$\pi \equiv {\pi_{\varphi}}^{\rho}=
-\frac{fH^3 (N^{\varphi})_{,\rho}}{2N^0}$ 
(with $\pi^{\rho \varphi}$ being the momentum conjugate to 
$g_{\rho \varphi}$), $E^{\rho}$ and $B$ are the electric and
magnetic fields and ${\cal{B}}$ is a boundary term.
The factor $\nu$ [defined in (\ref{T_Con_1})] comes from the fact 
that, due to the angle deficit, 
the integration over $\varphi$
is between $0$ and $2 \pi\nu$. 
Upon varying the action with respect to $f(\rho)$, $H(\rho)$, $\pi(\rho)$,
$\phi(\rho)$ and $E^{\rho}(\rho)$ one picks up additional surface terms.
Indeed,
\begin{eqnarray}
\delta S \hspace{-0.2cm}  &=& \hspace{-0.2cm} 
       - \Delta t N \nu{\biggl [}(H_{,\rho}-2H\phi_{,\rho})e^{-2\phi}
         \delta f^2 -(f^2)_{,\rho}e^{-2\phi}\delta H 
                                            \nonumber \\
 \hspace{-0.4cm}  & & \hspace{-0.4cm} 
         -4f^2 H e^{-2\phi} \delta(\phi_{,\rho})
          +2f^2 e^{-2\phi}\delta (H_{,\rho})
                                               \nonumber \\
 \hspace{-0.4cm}  & & \hspace{-0.4cm}       
     +2H{\bigl [}(f^2)_{,\rho}+4(\omega+1)f^2 \phi_{,\rho}{\bigr ]}
         e^{-2\phi}\delta \phi {\biggr ]}
                                                        \nonumber \\
  \hspace{-0.4cm}  & & \hspace{-0.4cm}
         +\Delta t N^{\varphi}\nu{\biggl [}2e^{-2\phi}\delta \pi
         -4 \pi e^{-2\phi}\delta \phi {\biggr ]}
                                                    \nonumber \\
  \hspace{-0.4cm}  & & \hspace{-0.4cm}
     + \Delta t A_t\nu {\biggl [}
             - \frac{4H}{f}e^{-2 \phi} \delta E^{\rho}{\biggr ]}            
         + \delta {\cal{B}}         
                                    \nonumber \\
  \hspace{-0.4cm}  & & \hspace{-0.4cm}        
         +(\mbox{terms vanishing when 
                    equations of motion hold}).
                                            \nonumber \\
                               \label{DELTA_ACCAO}
\end{eqnarray}
In order that the Hamilton's equations are satisfied,
the boundary term ${\cal{B}}$ has to be adjusted so that 
it cancels the above additional surface terms. 
More specifically one has
\begin{equation}
  \delta {\cal{B}} = -\Delta t N \delta M  +\Delta t N^{\varphi}\delta J
                  + \Delta t A_t \delta Q_{\rm e} \:,
                              \label{DELTA_B}
\end{equation}
where one identifies $M$ as the mass, $J$ as the angular momentum
 and $Q_{\rm e}$ as the electric
charge since they are the terms conjugate to the asymptotic
values of $N$, $N^{\varphi}$ and $A_t$, respectively.

To determine the mass, the angular momentum  and the electric
charge of the solutions one must take the spacetime that 
we have obtained and subtract the background
reference spacetime contribution, i.e., we choose the energy 
zero point in such a way that the mass, angular momentum and charge 
vanish when the matter is not present.

Now, note that for $\omega >-1$ (and $\omega <-2$), spacetime
(\ref{Met_1_J}) has an asymptotic metric given by
\begin{equation}
-{\biggl (}\gamma^2-\frac{\varpi^2}{\alpha^2} {\biggr )}
 \alpha^2 \rho^2 dt^2+ \frac{d \rho^2}{ \alpha^2 \rho^2}+
{\biggl (}\gamma^2-\frac{\varpi^2}{\alpha^2} {\biggr )}
 \rho^2 d \varphi^2 \:,
                                          \label{ANTI_SITTER}
\end{equation}
where $\gamma^2-\varpi^2 / \alpha^2=1$ so, it is asymptotically 
an anti-de Sitter spacetime.
For $\omega >-1$ (and $\omega <-2$) the anti-de Sitter spacetime is 
also the background reference spacetime, since the metric  
(\ref{Met_1_J}) reduces to (\ref{ANTI_SITTER}) if the
matter is not present ($b=0$ and $\chi_{\rm m}=0$).

Taking the subtraction of the background reference spacetime 
into account and noting that $\phi-\phi_{\rm ref}=0$ and that
$\phi_{,\rho}-\phi_{,\rho}^{\rm ref}=0$ we have that the mass, 
angular momentum and electric charge are given by
\begin{eqnarray}
M &=& \nu {\bigl [}(2H\phi_{,\rho}-H_{,\rho})e^{-2\phi}(f^2-f^2_{\rm ref})
                                                 \nonumber \\
 & &
       +(f^2)_{,\rho}e^{-2\phi}(H-H_{\rm ref})
      -2f^2 e^{-2\phi}(H_{,\rho}-H_{,\rho}^{\rm ref}) {\bigr ]}\:, 
                                               \nonumber \\
J &=&  -2\nu e^{-2\phi} (\pi-\pi_{\rm ref}) \:,  
                                               \nonumber \\
Q_{\rm e} &=&  \frac{4H}{f} \nu e^{-2 \phi} 
             (E^{\rho}-E^{\rho}_{\rm ref}) \:.         
                                    \label{MQ_GERAL}
\end{eqnarray}  
Then, we finally have
that the mass and angular momentum are (after taking the asymptotic limit, 
$\rho \rightarrow +\infty$)
\begin{eqnarray}
 M &=& b \nu{\biggl [}\gamma^2
      + \frac{\omega +2}{\omega +1} 
      \frac{\varpi^2}{\alpha^2}{\biggl ]}
      + {\rm Div_M}(\chi_{\rm m},\rho) \:,  
                                        \label{M} \\
J &=&  \frac{\gamma \varpi}{\alpha^2} b \nu \frac{2\omega+3} 
{\omega +1}+ {\rm Div_J}(\chi_{\rm m},\rho)\:,  
                                    \label{J}
\end{eqnarray}  
where ${\rm Div_M}(\chi_{\rm m},\rho)$  and
${\rm Div_J}(\chi_{\rm m},\rho)$ 
are terms proportional to the magnetic source $\chi_{\rm m}$ 
that diverge as $\rho \rightarrow +\infty$. The presence of these 
kind of  divergences in the mass and angular momentum is a usual 
feature present in 
charged solutions. They can be found for example in the 
electrically charged point source solution \cite{Deser_Maz}, in the
electric counterpart of the BTZ black hole \cite{BTZ_Q}, in the 
pure electric black holes of 3D 
Brans-Dicke action \cite{OscarLemos} and in the magnetic
counterpart of the BTZ solution \cite{OscarLemos_BTZ}. Following 
\cite{Deser_Maz,BTZ_Q} (see also
\cite{OscarLemos,OscarLemos_BTZ}) the divergences on the mass can be 
treated as follows. One considers a boundary of large radius $\rho_0$
involving the system. Then, one sums and subtracts 
${\rm Div_M}(\chi_{\rm m},\rho_0)$ to (\ref{M}) 
so that the mass (\ref{M}) is now written as 
\begin{equation}
M = M(\rho_0)+ [{\rm Div_M}(\chi_{\rm m},\rho)-
     {\rm Div_M}(\chi_{\rm m},\rho_0)] \:,
           \label{M0_0}
\end{equation}  
where $M(\rho_0)=M_0+{\rm Div_M}(\chi_{\rm m},\rho_0)$, i.e.,
\begin{equation}
M_0=M(\rho_0)-{\rm Div_M}(\chi_{\rm m},\rho_0)\:.
                       \label{M0_0_v2}
\end{equation}  
The term between brackets in (\ref{M0_0}) vanishes when 
$\rho \rightarrow \rho_0$. Then $M(\rho_0)$ is the energy within the 
radius $\rho_0$. The difference between $M(\rho_0)$ and 
$M_0$ is $-{\rm Div_M}(\chi_{\rm m},\rho_0)$ which is 
interpreted as the electromagnetic energy outside $\rho_0$ 
apart from an infinite constant which is absorbed in 
$M(\rho_0)$. The sum (\ref{M0_0_v2}) is then independent of 
$\rho_0$, finite and equal to the total mass.
In practice the treatment of the mass divergence 
amounts to forgetting about $\rho_0$ and take as zero the asymptotic 
limit: $\lim {\rm Div_M}(\chi_{\rm m},\rho)=0$. 

To handle the angular momentum divergence, one first notices 
that the asymptotic limit of the angular momentum per unit  mass 
$(J/M)$ is either zero or one, so the angular momentum diverges at 
a rate slower or equal to the rate of the mass divergence. 
The divergence on the angular momentum can then be treated 
in a similar way as the mass divergence. So, one can again 
consider a boundary of large radius $\rho_0$
involving the system. Following the procedure applied for 
the mass divergence one concludes that the divergent 
term $-{\rm Div_J}(\chi_{\rm m},\rho_0)$ can be interpreted as 
the electromagnetic angular momentum outside  $\rho_0$ up to an 
infinite constant that is absorbed in $J(\rho_0)$.

Hence, in practice the treatment of both the mass and angular 
divergences amounts to forgetting about $\rho_0$ and take as 
zero the asymptotic limits : 
$\lim {\rm Div_M}(\chi_{\rm m},\rho)=0$ and 
$\lim {\rm Div_J}(\chi_{\rm m},\rho)=0$ in (\ref{M}) and (\ref{J}).

Now, we calculate the electric charge of the solutions. 
To determine the electric field we must consider the 
projections of the Maxwell field on spatial hypersurfaces. 
The normal to such hypersurfaces is 
$n^{\nu}=(1/N^0,0,-N^{\varphi}/N^0)$ and the electric field is
given by $E^{\mu}=g^{\mu \sigma}F_{\sigma \nu}n^{\nu}$. 
Then, from (\ref{MQ_GERAL}), the electric charge is
\begin{equation}
 Q_{\rm e}=-\frac{4Hf}{N^0} \nu e^{-2 \phi}(\partial_{\rho}A_t-N^{\varphi}
    \partial_{\rho} A_{\varphi})=\frac{\varpi}{\alpha^2} \nu
     \chi_{\rm m}\:.
\label{CARGA}
\end{equation}
Note that the electric charge is proportional
to $\varpi \chi_{\rm m}$.
In section \ref{sec:Phys_Interp} we will propose a physical 
interpretation for the origin of the magnetic field 
source and discuss the result obtained in (\ref{CARGA}).

The mass, angular momentum and electric charge of the static 
solutions can be obtained by putting $\gamma=1$ and 
$\varpi=0$ on the above expressions [see (\ref{TRANSF_J})].

\subsubsection*{{\bf (ii)} Brans-Dicke theory with $\omega=\pm \infty$}
The mass, angular momentum  and electric
charge of the spacetime generated by a 
magnetic point source in 3D Einstein-Maxwell 
theory with $\Lambda<0$ have been calculated in 
\cite{OscarLemos_BTZ}.
Both the static and rotating solutions have negative mass
and there is an upper bound for the intensity 
of the magnetic source and for the value of the angular
momentum.
 
\subsubsection*{{\bf (iii)} Brans-Dicke theories with $-\infty<\omega<-2$}

The mass, angular momentum  and electric
charge of the $\omega<-2$  solutions are also given by
(\ref{M}), (\ref{J}) and (\ref{CARGA}), respectively.
If $b<0$ the factor $\nu$ is 
defined in (\ref{T_Con_1}) and if $b>0$ one has $\nu=1$.

\vskip 3mm

For $-2 \leq \omega \leq -1$ [cases {\bf (iv)}-{\bf (vii)}], the asymptotic 
and background reference spacetimes have a very peculiar form. 
In particular, they are not an anti-de Sitter spacetime. Therefore, 
there are no conserved quantities for these 
solutions.

\subsection{The rotating magnetic solution in final form}

For cases {\bf (i)} $-1<\omega<+\infty$, {\bf (ii)}  $\omega=\pm \infty$ and 
{\bf (iii)} $-\infty<\omega<-2$  we may cast the metric  in terms of 
$M$, $J$ and $Q_{\rm e}$.

\subsubsection*{{\bf (i)} Brans-Dicke theories with $-1<\omega<+\infty$}

Use of (\ref{M}) and (\ref{J}) allows us  
to solve a quadratic equation for 
$\gamma^2$ and $\varpi^2 / \alpha^2$. It gives two distinct 
sets of solutions
\begin{equation}
\gamma^2=\frac{M(2- \Omega)}{2\nu b} \:,\:\:\:
\:\:\:\: \frac{\varpi^2}{\alpha^2}= 
  \frac{\omega+1}{2(\omega+2)}\frac{M \Omega}{\nu b}\:, 
\label{DUAS}
\end{equation}

\begin{equation}
\gamma^2=\frac{M \Omega}{2\nu  b} \:,\:\:\:
\:\:\:\: \frac{\varpi^2}{\alpha^2}=
      \frac{\omega+1}{2(\omega+2)}\frac{M(2- \Omega)}{\nu b}\:, 
\label{DUAS_ERR}
\end{equation}
where we have defined a rotating parameter $\Omega$ as
\begin{equation}
\Omega \equiv 1- \sqrt{1-\frac{4(\omega+1)(\omega+2)}
{(2\omega+3)^2}\frac{J^2 \alpha^2}{M^2}} \:.
                    \label{OMEGA}
\end{equation}
When we take $J=0$ (which implies $\Omega=0$), 
(\ref{DUAS}) gives $\gamma \neq 0$ and $\varpi= 0$ while
(\ref{DUAS_ERR}) gives the nonphysical solution $\gamma=0$ and 
$\varpi \neq 0$ which does not reduce to the static original metric. 
Therefore we will study the solutions found from (\ref{DUAS}).

For $\omega>-1$ (and $\omega<-2$),
the condition that $\Omega$ remains real imposes 
a restriction on the allowed values of the 
angular momentum
\begin{equation}
|\alpha J|\leq \frac{|2\omega+3|M}{2\sqrt{(\omega+1)
(\omega+2)}} \:.
                    \label{Rest_OMEGA}
\end{equation}

The parameter $\Omega$ ranges between 
$0 \leq \Omega \leq 1$. 
 The condition $\gamma^2-\varpi^2/\alpha^2=1$ fixes 
the value of $b$ as a function of $M,\Omega,\chi_{\rm m}$,
\begin{eqnarray}
b &=& \frac{M}{\nu}
{\biggl [} 1 - \frac{2\omega+3}{2(\omega +2)} \Omega {\biggr ]}\:,
                                \label{b} 
\end{eqnarray}
where 
\begin{eqnarray}
\nu = \frac{2(\omega +1)(\alpha r_+)^{\frac{\omega +2}{\omega +1}}
+M{\biggl (}1- \frac{2\omega +3}{2(\omega +2)}\Omega {\biggr )} }
{2(\omega +1)(\alpha r_+)^{\frac{2\omega +3}{\omega +1}}+2k \chi_{\rm m}^2
(\alpha r_+)^{-\frac{1}{\omega +1}}} \:.
                                  \label{b_2} 
\end{eqnarray}

The gravitational field (\ref{Met_1_J}) generated by the rotating 
point source may now be cast in the form
\begin{widetext}
\begin{eqnarray}
 ds^2 &=& 
    -{\biggl [}\alpha^2 (\rho^2 + r_+^2) 
           -\frac{\omega +1}{2(\omega +2)} \frac{M \Omega /\nu }
{[\alpha^2 (\rho^2 + r_+^2)]^{\frac{1}{2(\omega+1)}}}
    +\frac{(\omega +1)^2}{8(\omega +2)}\frac{Q^2_{\rm e}/\nu^2}
    {[\alpha^2 (\rho^2 + r_+^2)]^{\frac{1}{\omega+1}}}
          {\biggr ]} dt^2
                                             \nonumber \\
      & &   
         -\frac{\omega +1}{2\omega +3}\frac{J}{\nu}{\biggl [}
         [\alpha^2 (\rho^2 + r_+^2)]^{-\frac{1}{2(\omega+1)}}
  -\frac{(\omega +1)Q^2_{\rm e}}{4M \Omega \nu}
    [\alpha^2 (\rho^2 + r_+^2)]^{\frac{1}{\omega+1}}  
        {\biggr ]} 2dt d\varphi   
                                                 \nonumber \\
      & &    
       + \frac{\rho^2/(\rho^2 + r_+^2)}
       { {\biggl [}  \alpha^2 (\rho^2 + r_+^2) +\frac{M}{\nu} 
      \frac{1-2(\omega+3)\Omega/[2(\omega+2)]}{[\alpha^2 
      (\rho^2 + r_+^2)]^{\frac{1}{2(\omega+1)}}}
           -\frac{k\chi_{\rm m}^2}
    {[\alpha^2 (\rho^2 + r_+^2)]^{\frac{1}{\omega+1}}} {\biggr ]}} 
          d\rho^2
                                              \nonumber \\
      & &  
 + \frac{1}{\alpha^2}{\biggl [}
     \alpha^2 (\rho^2 + r_+^2)
   +\frac{M(2-\Omega)/(2\nu)}
   {[\alpha^2 (\rho^2 + r_+^2)]^{\frac{1}{2(\omega+1)}}}
    -\frac{(\omega+2)(2-\Omega)}{2(\omega+2)-(2\omega+3)\Omega}
     \frac{k\chi_{\rm m}^2}
    {[\alpha^2 (\rho^2 + r_+^2)]^{\frac{1}{\omega+1}}} 
      {\biggr ]}   d\varphi^2 \:,  \nonumber \\
                                 \label{Met_MJQ}  
\end{eqnarray}
with $0\leq\rho<\infty$ and the electromagnetic field generated 
by the rotating point source can be written as 
\begin{eqnarray}
A=\frac{A(\rho)}{\sqrt{2(\omega+2)-(2\omega+3)\Omega}}{\biggl [}
      -\alpha\sqrt{(\omega+1)\Omega} \: dt 
+\sqrt{(\omega+2)(2-\Omega)} \: d \varphi{\biggr ]} \:,   
\label{A_Fim}  
\end{eqnarray}
with $A(\rho)$ defined in (\ref{A_J}). 

The static solution can be obtained by putting 
$\Omega=0$ (and thus $J=0$ and $Q_{\rm e}=0$) on the above expression 
[see (\ref{TRANSF_J})].

\end{widetext}
\subsubsection*{{\bf (ii)} Brans-Dicke theory with $\omega=\pm \infty$}

The spacetime generated by a rotating 
magnetic point source in 3D Einstein-Maxwell 
theory with $\Lambda<0$ is written as a function of its hairs
in \cite{OscarLemos_BTZ}.
 
\subsubsection*{{\bf (iii)} Brans-Dicke theories with $-\infty<\omega<-2$}

For this range of $\omega$, the gravitational and electromagnetic 
fields  generated by a rotating magnetic point source are also given by
(\ref{Met_MJQ}) and (\ref{A_Fim}).
If $b<0$ the factor $\nu$ is 
defined in (\ref{b_2}) and if $b>0$ one has $\nu=1$.

\subsection{Geodesic structure}

The geodesic structure of the rotating spacetime is similar to 
the static spacetime (see section \ref{sec:geod}), although there are 
now direct (corotating with $L>0$) and retrograde 
(counter-rotating with $L<0$) orbits.  
The most important result that spacetime is geodesically complete 
still holds for the stationary spacetime.

\section{\label{sec:Phys_Interp} PHYSICAL INTERPRETATION 
OF THE MAGNETIC SOURCE}

When we look back to the electric charge given in
(\ref{CARGA}), we see that it is zero when $\varpi=0$, i.e., when
the angular momentum $J$
of the spacetime vanishes. 
This is expected since in the static solution 
we have imposed that the electric field is zero ($F_{12}$ is 
the only non-null component of the Maxwell tensor). 

Still missing is a physical interpretation for the origin
of the magnetic field source.
The magnetic field source is not a Nielson-Oleson vortex solution 
since  we are working with
the Maxwell theory and not with an Abelian-Higgs model.
We might then think that the magnetic field is produced
by a Dirac point-like monopole. However, this is not also the
case since a Dirac monopole with strength $g_{\rm m}$ appears 
when one breaks the Bianchi identity \cite{MeloNeto}, yielding 
$\partial_{\mu} (\sqrt{-g} \tilde{F}^{\mu} e^{-2 \phi})=
g_{\rm m} \delta^2 (\vec{x})$  (where
$\tilde{F}^{\mu}=\epsilon^{\mu \nu \gamma}F_{\nu \gamma}/2$ is 
the dual of the Maxwell field strength), whereas in this work
we have that  $\partial_{\mu} (\sqrt{-g} \tilde{F}^{\mu} e^{-2 \phi})=0$.
Indeed,  we are clearly 
dealing with the Maxwell theory which satisfies Maxwell equations 
and the  Bianchi identity 
\begin{eqnarray}
& &  \frac{1}{\sqrt{-g}}\partial_{\nu}(\sqrt{-g}F^{\mu \nu} e^{-2 \phi})
        =\frac{\pi}{2}\frac{1}{\sqrt{-g}} j^{\mu} \:,
                                    \label{Max_j} \\
& &  \partial_{\mu} 
     (\sqrt{-g} \tilde{F}^{\mu} e^{-2 \phi})=0 \:,         
                                    \label{Max_bianchi}
\end{eqnarray}  
respectively. In (\ref{Max_j}) we have made use of the fact 
that the general 
relativistic current density is $1/\sqrt{-g}$ times the 
special relativistic current density 
$j^{\mu}=\sum q \delta^2(\vec{x}-\vec{x}_0)\dot{x}^{\mu}$.

We then propose that the magnetic field source can be interpreted as
composed by a system of two symmetric and superposed electric charges
(each with strength $q$). One of the electric charges is at rest with
positive charge (say), and the other is spinning with an angular
velocity $\dot{\varphi}_0$ and negative electric charge.  Clearly,
this system produces no electric field since the total electric charge
is zero and the magnetic field is produced by the angular electric
current.
To confirm our interpretation, we go back to Eq. (\ref{Max_j}).
In our solution, the only non-vanishing component of the 
Maxwell field is $F^{\varphi \rho}$ which implies that only 
$j^{\varphi}$ is not zero. According to our interpretation
one has $j^{\varphi}=q \delta^2(\vec{x}-\vec{x}_0)\dot{\varphi}$, 
which one inserts in Eq. (\ref{Max_j}). 
Finally, integrating over $\rho$ and $\varphi$ we have
\begin{equation}
 \chi_{\rm m} \propto q \dot{\varphi}_0 \:.
\label{Q_mag}
\end{equation} 
So, the magnetic source strength, $\chi_{\rm m}$, can be 
interpreted as an electric charge times its 
spinning velocity.

Looking again to the electric charge given in
(\ref{CARGA}),
one sees that after applying the 
rotation boost in the $t$-$\varphi$ 
plane to endow the initial static spacetime with
angular momentum, there appears  a net electric charge.
This result was already expected since now, besides the 
scalar magnetic field ($F_{\rho \varphi} \neq 0$), there is also an 
electric field ($F_{t \rho} \neq 0$) [see (\ref{A_Fim})].
A physical interpretation for the appearance of the net 
electric charge is now needed. To do so, we return 
to the static spacetime. 
In this static spacetime there is a static positive charge and 
a spinning negative charge of equal strength at the center. 
The net charge is then zero. Therefore,
an observer at rest ($S$) sees a density of positive charges 
at rest which is equal to the density of negative charges that are
spinning. Now, we perform a local rotational boost 
$t'= \gamma t- (\varpi/\alpha^2) \varphi$ 
and $\varphi' = \gamma \varphi-\varpi t\:$ 
to an observer ($S'$) in the static spacetime, so that
$S'$ is moving relatively to $S$.
This means that $S'$ sees a different charge density since a density
is a charge over an area and this area suffers a Lorentz contraction
in the direction of the boost.  Hence, the two sets of charge
distributions that had symmetric charge densities in the frame $S$
will not have charge densities with equal magnitude in the frame
$S'$. Consequently, the charge densities will not cancel each other in
the frame $S'$ and a net electric charge appears.  This was done
locally. When we turn into the global rotational Lorentz boost of
Eqs. (\ref{TRANSF_J}) this interpretation still holds.
The local analysis above is similar to the one that occurs when one
has a copper wire with an electric current and we apply a translation
Lorentz boost to the wire: first, there is only a magnetic field but,
after the Lorentz boost, one also has an electric field.  The
difference is that in the present situation the Lorentz boost is a
rotational one and not a translational one.

\section{CONCLUSIONS}

We have added the Maxwell term to the action of a generalized  
3D  dilaton gravity specified by the 
Brans-Dicke parameter $\omega$ 
introduced in \cite{Sa_Lemos_Static}. 
For the static spacetime the electric and magnetic fields cannot be 
simultaneously non-zero, i.e. there is no static dyonic solution.
Pure electrically charged solutions of the theory have been studied
in detail in \cite{OscarLemos}. 

In this work we have found geodesically complete
spacetimes generated by  static and rotating magnetic point sources.
These spacetimes are horizonless and many of them have a conical
singularity at the origin. 
These features are common in spacetimes generated by point 
sources in 3D gravity theories 
\cite{Jackiw_Review}-\cite{Brown_Hen}.
The static solution generates a scalar 
magnetic  field while the rotating solution produces, in addition, a 
radial electric field. 
The source for the magnetic field can be interpreted as
composed by a system of two symmetric and superposed electric 
charges. One of the electric charges is at rest and the other 
is spinning. This system produces no electric field since the 
total electric charge is zero and the scalar magnetic field is 
produced by the angular electric current. 
When we apply a rotational Lorentz boost to add angular
momentum to the spacetime, there appears a net electric charge
and a radial electric field.

For $\omega=\pm \infty$ our solution reduces to the magnetic
counterpart of the BTZ solution, i.e,
to the spacetime generated by a magnetic 
point source in 3D Einstein-Maxwell theory with 
$\Lambda<0$ analysed in 
\cite{CL1}, \cite{HW}-\cite{OscarLemos_BTZ}.
The solutions corresponding to the theories described by a 
Brans-Dicke parameter that belongs to the
range $-1<\omega<+\infty$ or $-\infty<\omega<-2$ have a
 behavior quite similar  
to the  magnetic counterpart of the BTZ solution.
For this range of the Brans-Dicke parameter, the solutions are 
asymptotically anti-de Sitter. This allowed us to calculate the 
mass, angular momentum and charge of the solutions. 
The presence of divergences at spatial infinity in the conserved 
quantities is an usual feature in electric solutions 
\cite{Deser_Maz,BTZ_Q,OscarLemos,OscarLemos_BTZ}.

The relation between spacetimes generated by point 
sources in 3D and 
cylindrically symmetric 4D solutions has been noticed by many 
authors 
(see e.g. \cite{Brown_book,DJH_flat,Cat,Lemos,OscarLemos_string}).
The $\omega=0$ solution considered in this paper is
the 3D counterpart of 
the 4D longitudinal magnetic string source
found in \cite{OscarLemos_string}.
Indeed, the  dimensional reduction of 4D general
relativity with one Killing vector yields 
the $\omega=0$ case of Brans-Dicke theory.
Further working relating 4D solutions
with 3D solutions
has been done in \cite{Z_L_2} where 
the two well known families of 4D black holes 
(i.e., the toroidal charged rotating anti-de Sitter black holes, 
and the Kerr-Newman-anti-de Sitter black holes) are analized  
and the interpretation of the fields and charges in terms of the  
three-dimensional point of view is given.

This paper closes our program of studying theories of the 
Brans-Dicke type in 3D initiated in
\cite{Lemos} and \cite{Sa_Lemos_Static} and continued in 
\cite{Sa_Lemos_Rotat} and \cite{OscarLemos}.


\begin{acknowledgments} 

This work was partially funded by Funda\c c\~ao para a Ci\^encia e
Tecnologia (FCT) through project CERN/FIS/43797/2001 and
PESO/PRO/2000/4014.  OJCD also acknowledges finantial support from the
portuguese FCT through PRAXIS XXI programme. JPSL thanks
Observat\'orio Nacional do Rio de Janeiro for hospitality.

\end{acknowledgments}

\appendix*
\section{General shape of  $e^{2\mu(r)}$ function}
In this appendix we study the radial dependence of the
 $e^{2\mu(r)}$ function defined in equations 
(\ref{MET_TODOS})-(\ref{MET-3/2}).
The qualitative shape of the $e^{2\mu(r)}$ function varies with the
Brans-Dicke parameter $\omega$ and with the mass parameter
$b$. However, 
we can form a small number of seven  cases for which the 
$e^{2\mu(r)}$ function and the spacetime structure has the 
same behavior. These seven cases are analysed in the text
and the corresponding figures describing the respective shape 
of the $e^{2\mu(r)}$ function are as follows, \newline
{\bf (i)} $-1<\omega<+\infty$ $\longrightarrow$  Fig. 1(a); \newline
{\bf (ii)} $\omega=\pm \infty$ $\longrightarrow$  Fig. 1(a); \newline
{\bf (iii)} $-\infty<\omega<-2$ :

  if $b>0$ $\longrightarrow$  Fig. 1(b), 

  if $b<0$ $\longrightarrow$  Fig. 1(c); \newline
{\bf (iv)} $\omega=-2$ $\longrightarrow$  Fig. 2(a); \newline
{\bf (v)}  $-2<\omega<-3/2$ :

  if $b>0$ $\longrightarrow$  Fig. 2(a), 

  if $b<0$ $\longrightarrow$  Fig. 2(b) or Fig. 2(c) depending on the 
         values; \newline 
{\bf (vi)} $\omega=-3/2$ $\longrightarrow$  Fig. 2(a) \newline
{\bf (vii)} $-3/2<\omega<-1$ $\longrightarrow$  Fig. 2(a).

From the analysis of Fig. 1 and Fig. 2, we see clearly that for the 
range ($\omega<-2$, $b>0$) the function  $e^{2\mu(r)}$ is always positive  
while for some values of the range ($-2<\omega<-3/2$, $b<0$) 
 the function  $e^{2\mu(r)}$ is always negative.
For the other ranges of ($\omega$, $b$), the $e^{2\mu(r)}$ function
can take positive or negative values depending on the value of the 
coordinate $r$. 
When $e^{2\mu(r)}$ is negative,
the spacetimes suffer an apparent change of signature 
from $+1$ to $-3$ (see section \ref{sec:stat_sol}).

\begin{figure}
\includegraphics{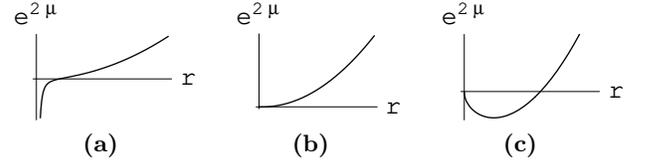}
{\small
\centerline{\noindent 
{\bf (a)} \hskip 2.2cm {\bf (b)} \hskip 2.2cm {\bf (c)} }
}
\caption{\label{Fig1}
General shape of $e^{2\mu(r)}$ for: 
{\bf (a)} Case (i)  $\omega>-1$ and 
case (ii) $\omega=\pm \infty$ ;
{\bf (b)}  Case (iii) $\omega<-2$ and $b>0$;
{\bf (c)} Case (iii) $\omega<-2$ and $b<0$.
}
\end{figure}

\begin{figure}
\includegraphics{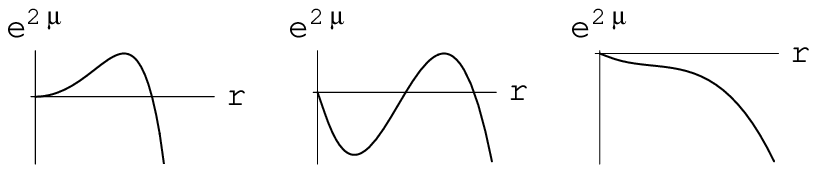}
{\small
\centerline{\noindent 
{\bf (a)} \hskip 2.2cm {\bf (b)} \hskip 2.2cm {\bf (c)} }
}
\caption{\label{Fig2}
General shape of $e^{2\mu(r)}$ for the
 values: 
{\bf (a)} (iv) $\omega=-2$, (v) $-2<\omega<-3/2$, $b>0$, 
(vi) $\omega=-3/2$, (vii) $-3/2<\omega<-1$;
{\bf (b)} some values of the range (v) ($-2<\omega<-3/2$, $b<0$);
{\bf (c)} some values of the range (v) ($-2<\omega<-3/2$, $b<0$).
}
\end{figure}


\end{document}